\documentclass[draft]{agujournal2019}
\journalname{The Planetary Science Journal}
\usepackage[utf8]{inputenc}

\usepackage{natbib}

\usepackage{url}
\usepackage{lineno}
\usepackage{soul}
\usepackage{amsmath}
\usepackage{appendix}
\usepackage{caption}
\usepackage{subcaption}
\usepackage{array}

\begin{document}
\justifying

   \title{Carbon cycling and habitability of massive Earth-like exoplanets}
   
\authors{Amanda Kruijver\affil{1,*}, Dennis H\"oning\affil{1,2,*}, Wim van Westrenen\affil{1}}

\affiliation{1}{Department of Earth Sciences, Vrije Universiteit Amsterdam, The Netherlands}
\affiliation{2}{Origins Center, The Netherlands}
\affiliation{*}{AK and DH equally contributed to this paper.}


\begin{abstract}
     As the number of detected rocky extrasolar planets increases, the question of whether their surfaces could be habitable is becoming more pertinent. On Earth, the long-term carbonate silicate cycle is able to regulate surface temperatures over timescales larger than one million years. Elevated temperatures enhance weathering, removing CO$_2$ from the atmosphere, which is subducted into the mantle. At mid-ocean ridges, CO$_2$ is supplied to the atmosphere from the interior. The carbon degassing flux is controlled by the melting depth beneath mid-ocean ridges and the spreading rate, influenced by the pressure- and temperature-dependent mantle viscosity. The influences of temperature and pressure on mantle degassing become increasingly important for more massive planets. Here, we couple a thermal evolution model of Earth-like planets of different masses with a model of the long-term carbon cycle and assess their surface temperature evolution. We find that the spreading rate at 4.5 Gyr increases with planetary mass up to 3 $M_\oplus$, since the temperature-dependence of viscosity dominates over its pressure-dependency. For higher mass planets, pressure-dependence dominates and the plates slow down. In addition, the effective melting depth at 4.5 Gyr as a function of planetary mass has its maximum at 3 $M_\oplus$. Altogether, at 4.5 Gyr, the degassing rate and therefore surface temperature have their maximum at 3 $M_\oplus$. This work emphasizes that both age and mass should be considered when predicting the habitability of exoplanets. Despite these effects, the long-term carbon cycle remains an effective mechanism that regulates the surface temperature of massive Earth-like planets.
\end{abstract}

\section{Introduction}
\label{intro}

Exoplanets are being discovered at a rapid rate. NASA's Kepler space telescope identified more than 4000 new planet candidates \citep{Fulton2017ThePlanets}, and advances in detection techniques are yielding observations of a vast array of distinct worlds \citep{Madhusudhan2016ExoplanetaryHabitability}. A significant fraction of currently known exoplanets has radii in between those of Earth and Neptune, $R_P$ = 1.0-3.9 $R_{\oplus}$ \citep{Batalha2013PlanetaryData}. The density of planets in this group with a radius smaller than 1.6 $R_{\oplus}$ is generally consistent with a rocky composition, as opposed to the larger planets having lower densities, seemingly suggesting a low-density envelope \citep{Fulton2017ThePlanets, Lozovsky2018ThresholdPlanets}. Although these small planets are strikingly common around Sun-like stars \citep{Marcy2014OccurrenceStars}, they have also been identified orbiting low mass stars. In the near future, the only properties of exoplanets that will be able to be explored beyond their ages, masses and radii are their atmospheres. For a few planets detailed information on their atmospheric compositions has been gathered and this will be extended to more super-Earths in the habitable zone in the near future by spectroscopic studies done by the ELT and the JWST \citep{Dorn2018OutgassingSuper-Earths}.

The prospect of finding conditions at which liquid water is stable at a planetary surface is exciting. Conditions that allow for liquid water would need to be present for extended periods of time for life to develop and to be sustained. A critical factor in the regulation of the surface climate is the long-term carbon cycle \citep{franck2000habitable, kadoya2014conditions, kadoya2015evolutionary, haqq2016limit,  Mills2019ModellingDay, Isson2020EvolutionEarth}.

Volcanic systems have vented CO$_2$ from Earth's mantle into the atmosphere for billions of years \citep{Kasting2003EvolutionPlanet}. Since silicate weathering and thereby the removal rate of CO$_2$ from the atmosphere depends on atmospheric CO$_2$ and surface temperature, a negative feedback is established. While these reactions have first been described by \cite{urey1952early}, the importance of the long-term carbon cycle as a feedback mechanism regulating the climate evolution of Earth and other planets has been described later \citep{Walker1981ATemperature,Kasting1989Long-termClimate}.

Carbon reaches the atmosphere from the deeper parts of the Earth through volcanism. At mid-ocean ridges, lithospheric plates diverge and mantle rock ascends to replace it. In the process, decompression melting takes place and carbon is degassed. The rate of carbon degassing is of fundamental importance as it influences the width of the habitable zone both for stagnant-lid \citep{Noack2017VolcanismZone} and plate tectonics planets \citep{kadoya2014conditions}. At the outer boundary, inefficient carbon degassing could lower the surface temperature to below the freezing point of water, while at the inner boundary high degassing rates could enhance surface temperatures and cause water evaporation \citep{Noack2017VolcanismZone}.

Degassing rates depend on the convective regime. While in the Solar System Earth is the only planet with active plate tectonics, the dominating tectonic regime on super-Earths is matter of debate \citep{Kite2009GeodynamicsPlanets}. \citet{Noack2017VolcanismZone} and \citet{Dorn2018OutgassingSuper-Earths} explore degassing rates for super-Earths in a stagnant-lid regime and find that degassing is most efficient on 2-3 $M_{\oplus}$ planets ($M_{\oplus}$ is the mass of Earth), depending on the initial mantle temperature of these planets. In this study, we explore the degassing flux and surface temperature for super-Earths with a mass between $M_\oplus$ = 1 and 10 with active plate tectonics. To this end, we couple a parameterized model of mantle convection to a model of the long-term carbon cycle. In contrast to previous carbon-cycle models, we model the degassing flux by including temperature- and pressure-dependent mantle viscosity and by explicitly calculating the melting depth beneath mid-ocean ridges.

\section{Model Set-Up}
\label{model}

Our model connects the long-term carbon cycle to the thermal evolution of plate tectonics planets of different masses. Carbon is degassed from the mantle reservoir $R_{man}$ into the atmospheric reservoir $R_{atm}$ by the degassing flux $F_{deg}$. Exposed weatherable rock reacts with atmospheric CO$_\text{2}$ to form bicarbonate and calcium and magnesium ions, which are washed into the oceans where they convert to carbonate minerals. An increase of CO$_\text{2}$ in the atmosphere speeds up silicate weathering, composed of the continental weathering flux $F_w$ and the seafloor weathering flux $F_{sfw}$, both of which draw CO$_\text{2}$ from the atmosphere. Carbon is added to the plate reservoir $R_p$ and subsequently transported by the subduction flux $F_{sub}$. A fraction of the subducted carbon is directly degassed back into the atmosphere at arc volcanoes ($F_{arc}$), while the rest enters the mantle, completing the long-term carbon cycle.

The basic carbon cycle model outlined above has been discussed extensively in the literature \citep{Walker1981ATemperature,Sleep2001CarbonEarth,Kasting2003EvolutionPlanet,Foley2015ThePlanets} and the main equations are given in \ref{carboncycle}. Here, we focus on the degassing rate and its evolution through time for different planetary masses. Degassing at mid-ocean ridges is the dominant pathway for carbon to reach the surface from the mantle \citep{orcutt2019deep} and is therefore crucial for the long-term carbon cycle. In previous studies, the degassing flux has often been taken as a predefined input flux when considering Earth's evolution \citep{krissansen2017constraining,kadoya2020probable} or taken to be controlled by the mantle carbon reservoir when considering the evolution of Earth-sized planets \citep{Sleep2001CarbonEarth,honing2019carbon}. \cite{Foley2015ThePlanets} introduced a dependence of the plate velocity - and thereby degassing rate - on the surface temperature, but found that the effect on the degassing rate for planets in the habitable zone is small. \cite{oosterloo2021} calculated the plate velocity as a function of the mantle heat flow but kept the melting depth constant and did not account for the pressure-dependence of mantle viscosity, making an application to more massive planets challenging. \cite{kadoya2015evolutionary} also neglected the pressure-dependence of viscosity when considering the carbon cycle for exoplanets.

Following \cite{Sleep2001CarbonEarth} and \cite{Foley2015ThePlanets}, the degassing flux $F_{deg}$ is given by
\begin{linenomath}
\begin{equation}
    F_{deg} = f_d \frac{R_{man}}{V_{m}} \ 2v_p L d_{melt},
\end{equation}
\label{eq:fdeg}
\end{linenomath}
where $f_d$ is the fraction of upwelling mantle that degasses, $v_p$ the plate velocity, $L$ length of ridges and $d_{melt}$ the depth where melting begins. The carbon density of a planet is given by the ratio between the mantle carbon reservoir $R_{man}$ and the mantle volume $V_m$. In the following, we will focus on the parameters $v_p$ and $d_{melt}$, which  differ substantially as a function of planet mass and thereby control the surface habitability.

With increasing planetary mass, the pressure increases, leading to an increase of the mantle viscosity. According to \citet{Tackley2013MantleViscosity}, the core-mantle boundary (CMB) region on a 10 $M_{\oplus}$ planet with an Earth-like composition experiences a ten times pressure increase. This significant pressure increase substantially slows down mantle convection \citep{Miyagoshi2013OnSuper-Earths}. \citet{vandenBerg2019Mass-dependentProperties} show that the effect of increasing mass on viscosity could result in a viscosity of two orders of magnitude higher for a 8 $M_\oplus$ than for an Earth-sized planet. This increase in viscosity for higher mass planets is further supported by \citet{Schaefer2015TheCycle}, who show that a $M$ = 5 planet develops a mantle viscosity up to two orders of magnitude higher when a pressure-dependent viscosity case is considered as opposed to a pressure-independent viscosity, underscoring the importance of including the pressure effect on viscosity. However, higher-mass planets do not solely imply an increase in internal pressures, but also exhibit higher internal temperatures. For example, retention of primordial heat could be higher due to a relatively smaller ratio of surface area to planetary volume or due to ineffective convective heat transport caused by the higher pressure affecting viscosity.

We base out interior thermal evolution model on \cite{Schaefer2015TheCycle} and derive the plate velocity in Eq. \ref{eq:fdeg} from boundary layer theory \citep{Schubert2001MantlePlanets}
\begin{linenomath}
\begin{equation}
\label{convvel}
	v_p =  \frac{5.38\kappa(R_p-R_c)}{{\delta_u}^2},
\end{equation}
\end{linenomath}
where $\kappa$ is the mantle thermal diffusivity, $R_p$ and $R_c$ are the radius of the planet and core respectively, and $\delta_u$ is the upper boundary layer thickness, given by
\begin{linenomath}
\begin{equation}
\label{ubl}
	\delta_u = D\left(\frac{Ra_{crit}}{Ra}\right)^\beta = \left(\frac{\kappa\eta(\langle T_m \rangle,P_{mm})Ra_{crit}}{g\alpha\rho_m\Delta T}\right)^\beta,
\end{equation}
\end{linenomath}
where $D$ is the mantle thickness, $\eta(\langle T_m \rangle,P_{mm})$ is the effective mantle viscosity with the mid-mantle pressure $P_{mm}$ and the spherically averaged mantle temperature $\langle T_m \rangle$, $\alpha$ the thermal expansivity, $\Delta T$ is the super-adiabatic temperature increase throughout the mantle, and $Ra$ and $Ra_{crit}$ are the Rayleigh and critical Rayleigh number, respectively.

Since in contrast to \cite{Schaefer2015TheCycle} we do not model the deep water cycle, we calculate the viscosity for dry silicate material following \cite{stamenkovic2011thermal,stamenkovic2012influence}. The effective viscosity is then given by

\begin{linenomath}
\begin{equation}
\label{viscosity}
    \eta(\langle T_m \rangle,P_{mm}) = \eta_{0} \exp\left\lbrace\frac{E_a}{R_{gas}}\left(\frac{1}{\langle T_m \rangle}-\frac{1}{T_{ref}}\right)+\frac{1}{R_{gas}}\left(\frac{P_{mm}V_{a}}{\langle T_m \rangle}-\frac{P_{ref}V_{a,ref}}{T_{ref}}\right)\right\rbrace,
\end{equation}
\end{linenomath}

where the values for the reference temperature and pressure are taken to be 1600 K and 0 Pa, respectively, and the pressure-dependent activation volume is given by the scaling law derived by \citet{stamenkovic2011thermal}

\begin{linenomath}
\begin{equation}
\label{activationvolume}
   V_a=\left(1.38+2.15\cdot \exp\left[-0.065\left(\frac{P_{mm}}{10^9}+10\right)^{0.485}\right]\right)\cdot 10^{-6}.
\end{equation}
\end{linenomath}

The average temperature follows from an adiabatic temperature profile extrapolated to the surface \citep{Schaefer2015TheCycle}
\begin{linenomath}
\begin{equation}
    \langle T_m \rangle = \frac{3}{{R_p}^3-{R_c}^3} \int_{R_c}^{R_p} r^2T_{ad}(r)dr,
\end{equation}
\end{linenomath}
with the adiabatic temperature profile
\begin{linenomath}
\begin{equation}
\label{adiabat}
    T_{ad}(r) = T_p + T_p\frac{\alpha g}{c_p}\Delta r,
\end{equation}
\end{linenomath}
where $T_p$ is the potential temperature of the mantle.

We calculate the thickness of the lower thermal boundary layer similar to the upper thermal boundary layer (Eq. \ref{ubl}) but use a local instability criterion, since the high pressure at the core-mantle boundary substantially influences the heat flow here. This calculation differs from the calculation of the upper boundary layer thickness (Eq. \ref{ubl}), which depends on the convection rate of the whole mantle, and therefore requires the use of an effective viscosity.

The thickness of the lower thermal boundary layer $\delta_l$ can then be calculated as

\begin{linenomath}
\begin{equation}
\label{lbl}
    \delta_l = \left(\frac{\kappa\eta(T_c,P_c)Ra_{crit}}{g\alpha\rho_m(T_c-T_l)}\right)^\beta,
\end{equation}
\end{linenomath}

where $T_c$ and $P_c$ are the temperature and pressure at the core-mantle boundary and $T_l$ is the temperature at the top of the lower thermal boundary layer.

The initial core temperature $T_{c,i}$ is estimated using a scaling law derived by \citet{Stixrude2014MeltingSuper-earths}

\begin{linenomath}
\begin{equation}
    T_{c,i} = T_{c,i\oplus} \left(\frac{P_c}{140}\right)^{0.48},
\end{equation}
\end{linenomath}

where $T_{c,i\oplus} =$ 4180 K \citep{fiquet2010melting}. Following \cite{Schaefer2015TheCycle}, we assume an initial mantle potential temperature of 2520 K for all planetary masses in our reference model, which translates into an initial average mantle temperature of 3000 K for the Earth, increasing with planetary mass. The effect of assuming different initial mantle potential temperatures is assessed separately (Section \ref{carbondensity}). Calculating the heat fluxes out of the core and mantle, we apply a standard parameterized thermal evolution model from \cite{Schubert2001MantlePlanets}, which is described in \ref{thermalevolution}.

\begin{figure}
 \centering 
 \includegraphics[scale=0.23]{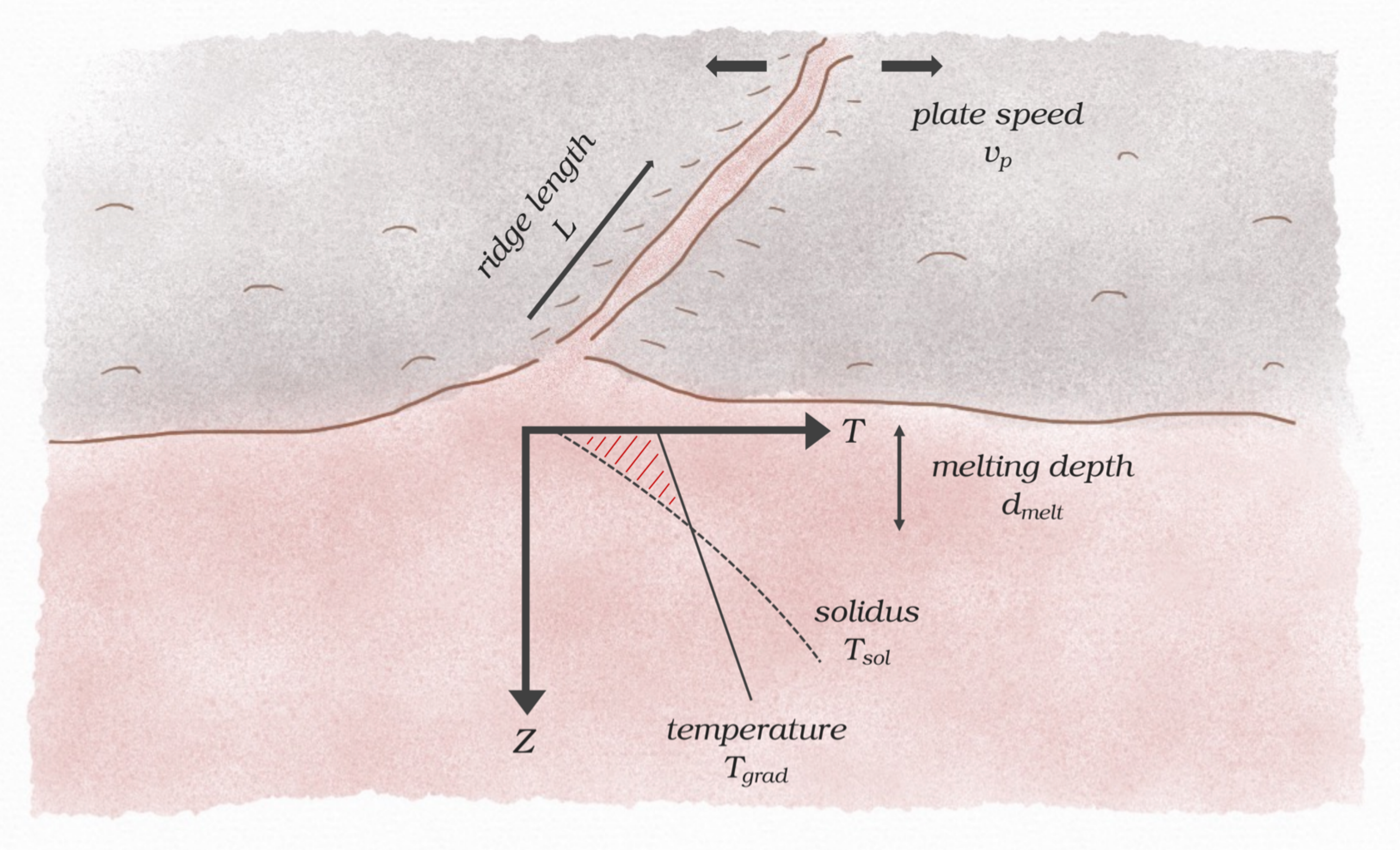}
 \caption{The melting region at mid-ocean ridges is determined by the depth at which melting begins $d_{melt}$. The production rate of oceanic crust additionally depends on  the plate speed $v_p$ and the ridge length $L$.}
 \label{meltdraw}
\end{figure}

In addition to the plate velocity, the mantle melting depth must be calculated in order to determine the degassing flux. Since we consider melting at mid-ocean ridges, we follow \cite{Kite2009GeodynamicsPlanets} and assume that melting occurs as long as the mantle potential temperature is higher than the zero-pressure solidus. The melting depth is considered to be the region between the surface and the point where the solidus crosses the adiabatic temperature gradient ($T_{sol} > T_{grad}$), as seen in figure \ref{meltdraw}. Since with increasing planetary mass planets are expected to cool more slowly or even heat up, the melting depth increases with time relative to 1 Earth mass planets \citep{Papuc2008ThePlanets}.

The solidus curve for pressures up to 10 GPa is given by \citet{Hirschmann2000MantleComposition}
\begin{linenomath}
\begin{equation} \label{eq:Tsol1}
    T_{sol} = 1397.273+137.863 P - 5.722 P^2,
\end{equation}
\end{linenomath}
and for pressures between 10 and 12 GPa, the solidus temperature is given by a later work from the same authors \citep{Hirschmann2009DehydrationPartitioning}
\begin{linenomath}
\begin{equation} 
\label{eq:Tsol2}
    T_{sol} = -1.092(P-10)^2 + 32.39(P-10) + 2208.15.
\end{equation}
\end{linenomath}

For pressures higher than 12 GPa, the solidus temperature becomes irrelevant, since any produced melt is likely unable to rise to the surface: The P-T location of crystal-liquid density inversions is determined by the compressibility, phase equilibria and element partitioning \citep{agee1998crystal}. At high pressures, density inversions are most likely if the liquid has a high compressibility relative to the crystalline phase, when the crystalline phase is stable over a large pressure range or when the liquid is 'chemically dense' \citep{agee1998crystal}. According to \citet{agee1998crystal}, crystal-liquid density inversions occur in the Earth in the transition zone and lower mantle, due to the extensive stability range of phases such as garnet and perovskite. For simplicity, we follow \citet{Noack2017VolcanismZone} and use a fixed value of 12 GPa as estimated by \citet{agee1998crystal} and \citet{ohtani1995melting}. In our model, we impose a limit to the effective melting depth at this pressure.

As noted above, the remaining carbon fluxes are calculated following standard carbon cycle models \citep{Sleep2001CarbonEarth,Foley2015ThePlanets} and given in \ref{carboncycle}. All parameters that directly depend on the size and/or mass of the planet (such as subduction zone length, land area, and gravity) are adjusted accordingly. For Earth-sized planets, the initial carbon distribution between the atmosphere, crust, and mantle has been shown to become unimportant after 1 Gyr \citep{Foley2015ThePlanets,honing2019carbon}. For simplicity, we assume for our reference model that all carbon is initially stored in the mantle, but will elaborate on the effect of the initial carbon distribution in Section \ref{carbondensity}. We also do not explicitly model ocean chemistry, which only affects the climate on considerably shorter timescales, since the carbonate-silicate cycle takes over control of the system on timescales longer than $\approx$1 Myr \citep{sundquist1991steady,colbourn2015time,honing2020impact}.

\cite{Valencia2006InternalPlanets,valencia2007Inevitability} calculate the interior structure of super-Earths up to 10 Earth masses $M_\oplus$. The authors find that the dependence of several parameters on mass can be adequately described by a power-law relationship of the form:
\begin{linenomath}
\begin{equation}
\label{powerlaw}
    Z = Z_\oplus \left(\frac{M}{M_\oplus}\right)^\xi,
\end{equation}
\end{linenomath}
where $Z$ is the considered parameter and $\xi$ is a scaling exponent. In this way parameters such as gravitational acceleration, mantle density and planetary radius can be related to planet mass. For example, we use the following power law for the mantle density $\rho_m$:
\begin{linenomath}
\begin{equation}
\label{density}
\rho_m = \rho_{m \oplus} \bigg(\frac{M}{M_{\oplus}}\bigg)^{0.195}.
\end{equation}
\end{linenomath}
Values for $\xi$ for the other scaled parameters can be found in Table \ref{tab:2}.


\begin{table}
\centering
 \caption{Parameters used in this study. References: $^1$\cite{Turcotte2002Geodynamics}, $^2$\cite{Foley2015ThePlanets}, $^3$\cite{Schaefer2015TheCycle}, $^4$\cite{Hirschmann2000MantleComposition} and $^5$\cite{hirschmann2018comparative}}
 \begin{tabular}{l l l l} 
 \textbf{Parameter} &  & \textbf{Value} & \textbf{Reference} \\ [0.5ex] 
 \hline
 $M_{\oplus}$ & mass Earth & $5.9736\cdot10^{24}$ kg & $^1$ \\ 
 $f_d$ & fraction mantle that degasses & 0.32 & $^2$ \\ 
 $\kappa$ & mantle thermal diffusivity & $10^{-6}$ m$^2$ s$^{-1}$ & $^3$ \\
 $Ra_{crit}$ & critical Rayleigh number & 1100 & $^3$  \\
 $\alpha$ & thermal expansivity & $2\cdot 10^{-5}$ K$^{-1}$ & $^3$  \\
 $\beta$ & scaling exponent & 1/3 & $^3$  \\ 
 $\eta_0$ & viscosity & $10^{21}$ Pa s & $^3$ \\
 $E_a$ & activation energy & 335 kJ mol$^{-1}$ & $^3$  \\
 $E_{a,lm}$ & activation energy lower mantle & 300 kJ mol$^{-1}$ & $^3$ \\
 $V_a$ & activation volume & 4 cm$^3$ mol$^{-1}$ & $^3$ \\
 $V_{a,lm}$ & activation volume lower mantle & 2.5 cm$^3$ mol$^{-1}$ & $^3$ \\
 $C_{tot}$ & Bulk silicate Earth carbon budget & $5.56 \cdot 10^{20}$ kg & $^5$ \\
 $C_p$ & mantle heat capacity & 1200 J kg$^{-1}$ K$^{-1}$ & $^3$   \\
 $C_{p,c}$ & core heat capacity & 840 J kg$^{-1}$ K$^{-1}$ & $^3$  \\
 $R_{gas}$ & ideal gas constant & 8.314 J mol$^{-1}$ K$^{-1}$ & $^3$ \\
 $P_{ref}$ & reference pressure & 0 & $^3$ \\
 $T_{ref}$ & reference temperature & 1600 K & $^3$ \\
 $T_{p,i}$ & initial mantle potential temperature & 2520 K & $^3$ \\
 $k$ & mantle thermal conductivity & 4.2 W m$^{-1}$ K$^{-1}$ & $^3$  \\ 
 $A_1$ & constant & 1397.273 K & $^4$  \\
 $A_2$ & constant & 137.863 K & $^4$ \\
 $A_3$ & constant & -5.722 & $^4$ \\ 
 \hline
  \label{tab:1}
 \end{tabular}

\end{table}

\begin{table}
\centering
\caption{Parameters that are scaled according to mass, from \cite{Valencia2006InternalPlanets}}

 \begin{tabular}{l l l l} 
 \hline
 Parameter & Description & Baseline value & $\xi$ \\ 
 \hline
 $R$ & Planetary radius & 6371 km & 0.269\\
 $R_c$ & Core radius & 3480 km & 0.247\\
 $g$ & Gravitational acceleration & 9.81 km &0.462\\
 $\rho_m$ & Mantle density & 3300 kg m$^{-3}$ &0.195\\
 $R_{man}$ & Bulk silicate Earth carbon budget & $5.56 \cdot 10^{20}$kg & 1\\
 $L$ & Subduction zone length & $6 \cdot 10^7$ m & 0.269\\
 \hline
   \label{tab:2}
\end{tabular}
\end{table}

\section{Results}
We first focus on the effective melting depth at 4.5 Gyr for planets with masses between 1 and 10 Earth masses (Section \ref{stablestateT}). In Section \ref{climateevol}, we elaborate on the interior and surface temperature evolution, and Section \ref{carbondensity} addresses the effect of different initial mantle temperatures and carbon concentrations.

\subsection{Effective Melting Depth and Plate Thickness at 4.5 Gyr}
\label{stablestateT}

\begin{figure}
    \centering
    \includegraphics[width=\textwidth]{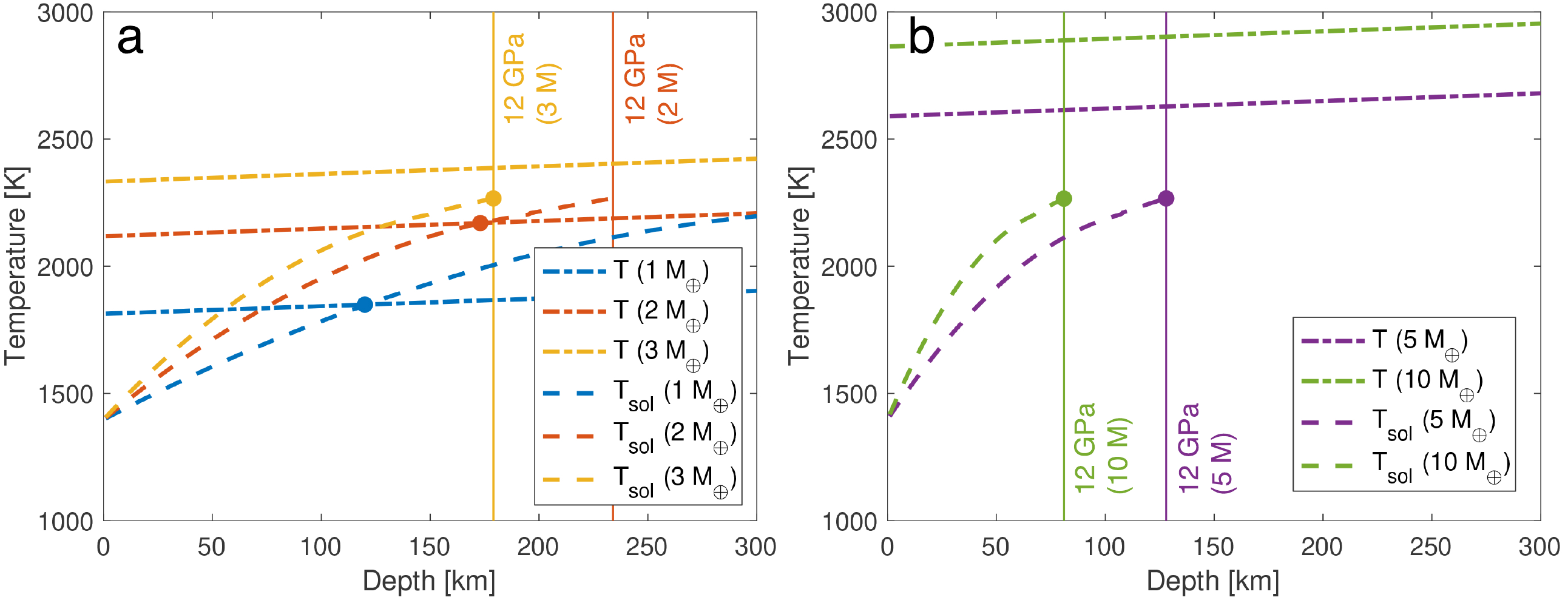}
    \caption{Temperature profile adiabatically extrapolated to the surface after 4.5 Gyr of the evolution (dashed-doted) and solidus temperatures (dashed) for different planet masses. The thin solid lines correspond to the depth below which melt is negatively buoyant. The filled circles correspond to the effective melting depth, which is either the intersection between the temperature profile and the solidus temperature or the depths below which melt is negatively buoyant.}
    \label{fig:meltregion}
\end{figure}

The melting depth is a main parameter controlling the surface temperature in our model as it directly affects the degassing rate. With increasing planetary mass, both the mantle temperature at 4.5 Gyr and the solidus temperature increase, which are competing factors determining the melting depth. In addition, the 12 GPa limit to positively buoyant melt moves to shallower depth. This limit is imposed to account for the densification of magmatic liquids up to the point where the melt will become neutrally or even negatively buoyant with respect to co-existing minerals \citep{agee1998crystal, ohtani1995melting}. All three parameters for different planetary masses are illustrated in Fig. \ref{fig:meltregion}, where the effective melting depth is marked. It becomes apparent that as long as the melting depth is reached for pressures below the 12 GPa limit, the melting depth at 4.5 Gyr steadily increases with planetary mass. In contrast, for planets whose effective melting depth is determined by the 12 GPa limit, this depth decreases with increasing planetary mass. Therefore, a peak in the melting depth for intermediate planetary masses is expected.

\begin{figure}
    \centering
    \includegraphics[width=\textwidth]{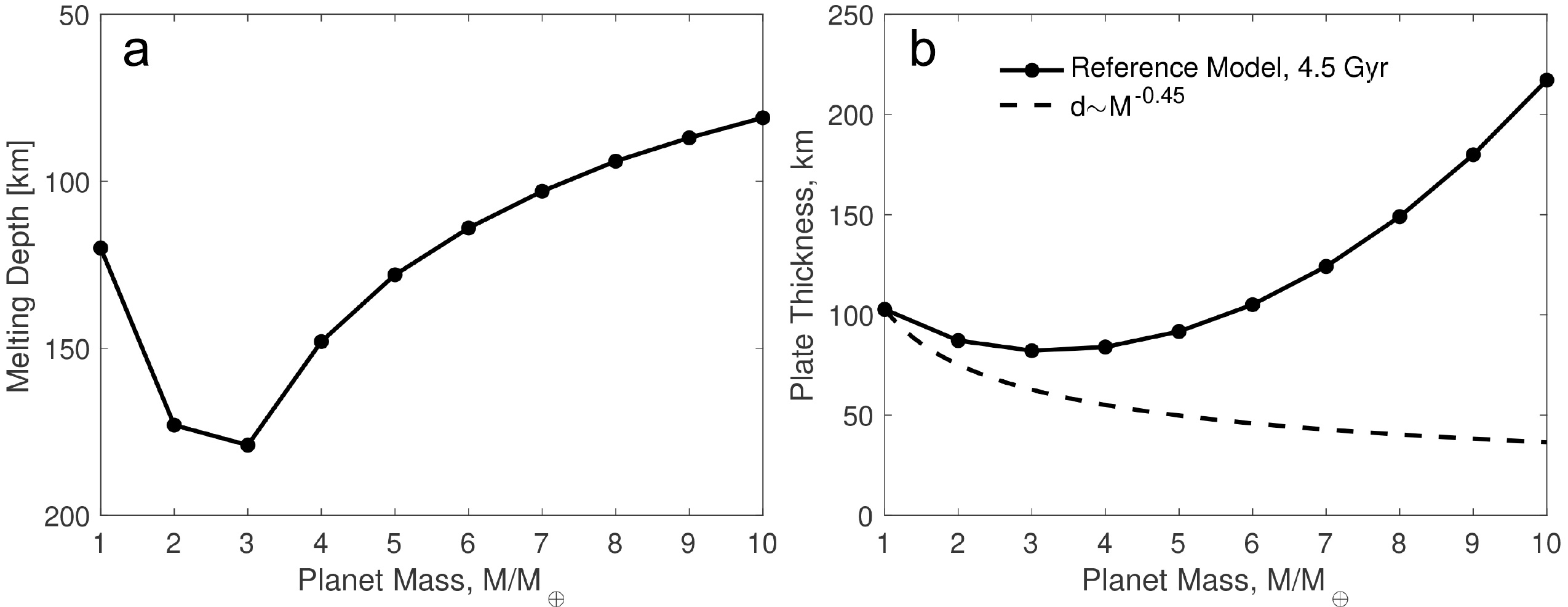}
    \caption{(a) Melting depth and (b) plate thickness at 4.5 Gyr as a function of planetary mass (dashed: Scaling extrapolation neglecting pressure-dependent viscosity after \citealt{valencia2007Inevitability}).}
    \label{fig:meltdepth}
\end{figure}

In Fig. \ref{fig:meltdepth}, we show (a) the effective melting depth and (b) the plate thickness at 4.5 Gyr as a function of planetary mass. For the effective melting depth, we find a minimum at 3 Earth masses: On the one hand, the mantle temperature generally increases stronger with planetary mass than the solidus temperature does, resulting in an increase of the melting depth. On the other hand, the 12 GPa limit moves shallower to shallower depths, which becomes dominant for planets from 3 Earth masses. As a result, the effective melting depth, i.e. the depth that controls degassing, has its peak at 3 Earth masses (see Fig. \ref{fig:meltdepth}a).

For the plate thickness (Fig. \ref{fig:meltdepth}b), we also find a minimum at 3 Earth masses. This is a result of the pressure-dependence of viscosity, which dominates over its temperature-dependence for planets more massive than 3 $M_{\oplus}$. Since the plate velocity directly follows from the boundary layer thickness (see Eq. \ref{convvel}), this effect also promotes high degassing rates particularly for planets of 3 Earth masses. We note that this minimum does not exist for a model that neglects the pressure-dependence of viscosity (dashed curve, following \citealt{valencia2007Inevitability}). Altogether, accounting for pressure-dependent viscosity, both the plate thickness and the melting depth promote a peak of the degassing rate at 4.5 Gyr for planets of $\approx3M_{\oplus}$.

\subsection{Climate Evolution for Different Planetary Masses}
\label{climateevol}

\begin{figure}
    \centering
    \includegraphics[width=\textwidth]{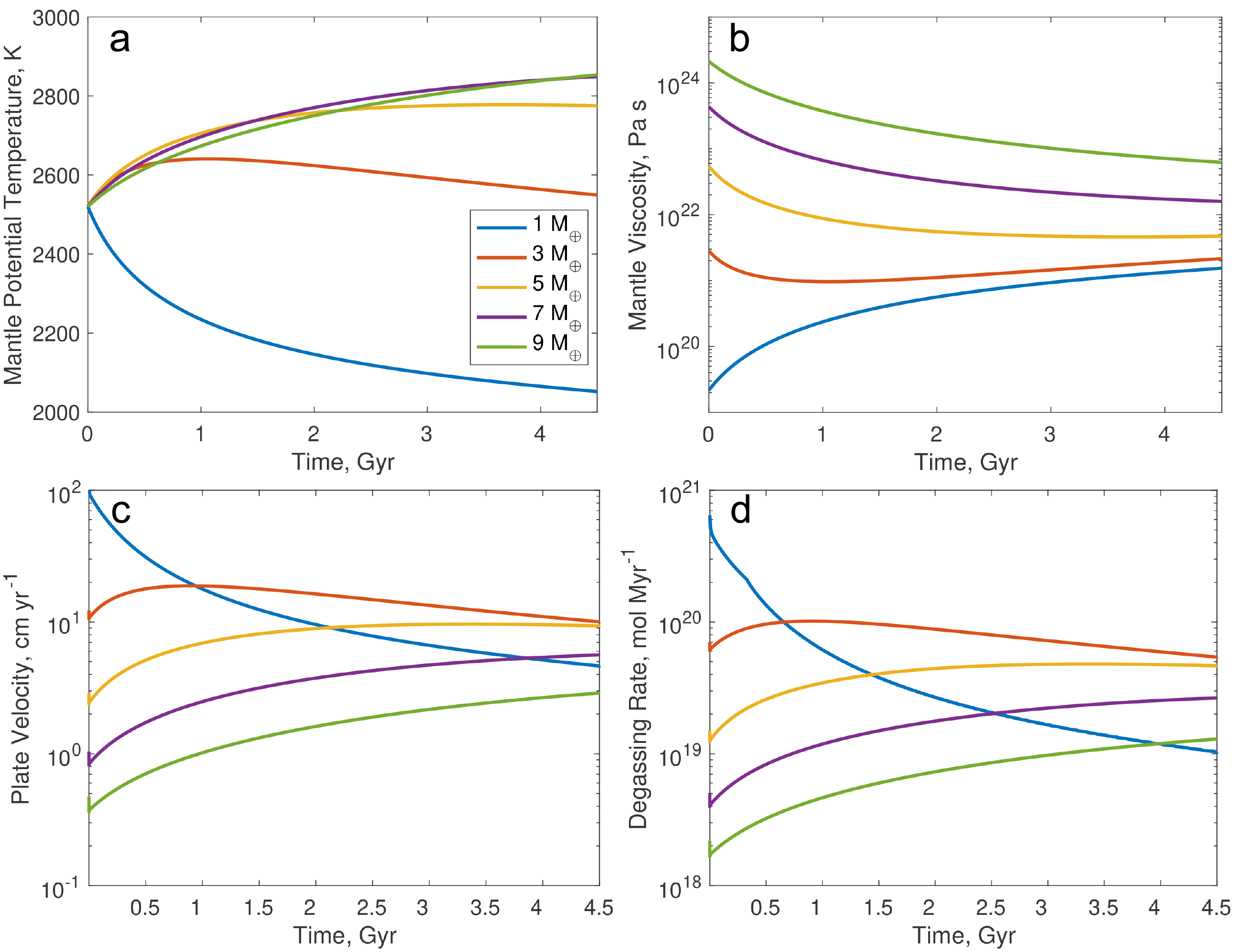}
    \caption{Evolution of (a) mantle potential temperature, (b) mantle viscosity,} (c) plate velocity and (b) degassing rate for different planet masses.
    \label{fig:degassing}
\end{figure}

Increasing planetary mass generally implies a delayed cooling of the mantle or even an increasing temperature with time due the higher mantle viscosity. As a consequence, more massive planets remain hotter for a longer period of time (Fig. \ref{fig:degassing}a). However, the plate velocity (Fig. \ref{fig:degassing}c) depends on the mantle viscosity and therefore not only on temperature but also on pressure. From $\approx$1 to 4.5 Gyr, this combination results in a particularly high plate velocity for planets with 3 Earth masses. This result, however, depends on the initial mantle temperature (discussed later, Section \ref{carbondensity}). Since the degassing rate (Fig. \ref{fig:degassing}d) depends on both the plate velocity and the melting depth, its evolution is similar to that of the plate velocity, with an even stronger decrease of the degassing rate with time for 1 Earth mass planets compared to more massive planets. This is a consequence of the more rapid cooling of 1 Earth mass planets, which  strongly reduces the melting depth with time.

\begin{figure}
    \centering
    \includegraphics[width=\textwidth]{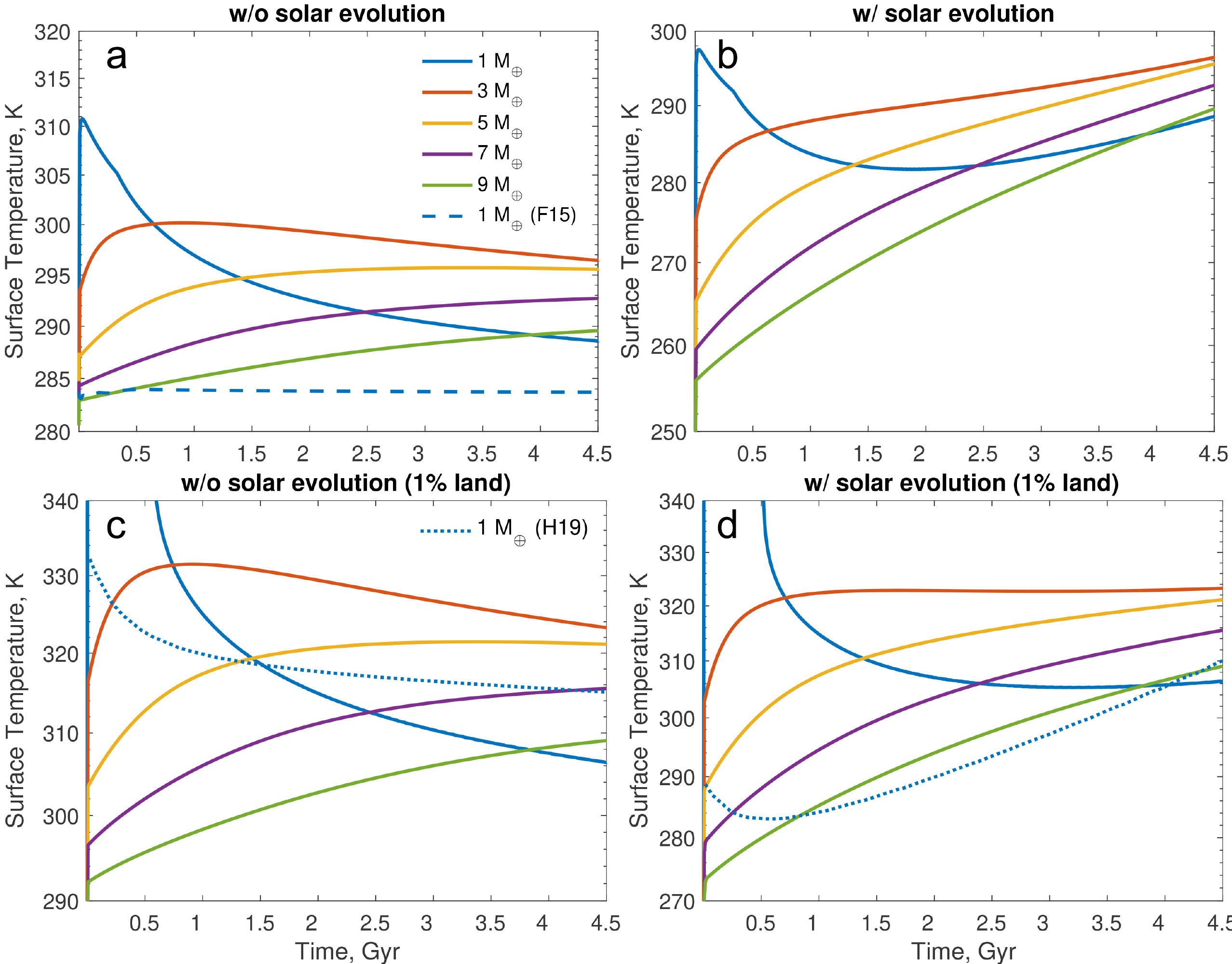}
    \caption{Surface temperature evolution without (left) and with (right) solar evolution for our reference model of 30\% land are (top) and a reduced land area of 1\% (bottom). The dashed line in (a) corresponds to the results of \cite{Foley2015ThePlanets} and the dotted lines in (c) and (d) to the results of \cite{honing2019carbon}.}
    \label{fig:surfT}
\end{figure}

The evolution of the surface temperature for different planetary masses is depicted in Fig. \ref{fig:surfT} for models with (a and c) and without (b and d) solar evolution. For the models without solar evolution (panels b and d), we use the present-day solar luminosity, and for the models that consider solar evolution (panels a and c), the luminosity is assumed to increase linearly by 1/3 during 4.5 Gyr up to the present-day value \citep{ribas2009sun}. Other orbital and stellar parameters are kept at Earth's values. While panels (a) and (b) depict modelling results for an Earth-like planet with an emerged land fraction of 30\%, panels (c) and (d) depict planets where the land fraction is reduced to 1\%. For comparison with previous carbon cycle studies for 1 Earth mass planets, modelling results from \cite{Foley2015ThePlanets} are added to panel (a) and modelling results from \cite{honing2019carbon} are added to panels (c) and (d).

Since \cite{Foley2015ThePlanets} neglected the effect of the interior evolution on the degassing rate, the temperature evolution in their model stays constant after an initial adjustment time (dashed curve in Fig. \ref{fig:surfT}a). In contrast, our results indicate a steadily decreasing surface temperature for 1 Earth mass planets (solid blue in panel a) if solar evolution is neglected, since both the plate velocity and the melting depth decrease with time. The fact that both curves do not cross at 4.5 Gyr is because we do not scale the present-day degassing rate to match a predefined value. For intermediate planet masses (red and yellow curves in panel a), the surface temperature does not substantially vary over time, and for high-mass planets (purple and yellow curves), the surface temperature steadily increases. If solar evolution is considered, the low- and intermediate mass planets in Fig. \ref{fig:surfT}b have the smallest temperature variation throughout the past 3 Gyr, as the mantle temperature decreases with time (or, at least, does not significantly increase). We also find that the most massive planets have a surface temperature below the freezing point of water in their early evolution, (discussed in more detail in Section \ref{discussion}).

Considering a planet with only 1\% emerged land area (Fig. \ref{fig:surfT}c, d), the differences between planets of different masses increase, since climate regulation does not work as efficiently in the absence of a substantial contribution of temperature-dependent continental weathering. However, the trend found in Fig. \ref{fig:surfT} (a and b) for the surface temperature evolution for different planetary masses remains the same. For comparison, the dotted curve shows modelling results from \cite{honing2019carbon}, considering a carbon cycle with a temperature-dependent arc-volcanism flux for water-covered planets of one Earth mass. While both modelling results suggest a decreasing surface temperature if solar evolution is neglected, the surface temperature in our model with a degassing rate dependent on plate velocity and melting depth depends on the thermal evolution more strongly. We also find that the high initial degassing rate for 1 Earth mass planets in combination with the limited weathering rate owed to the small land fraction imply hot, potentially uninhabitable surface conditions during the first 0.5 Gyr.

\subsection{Initial Mantle Temperature and Carbon Budget} \label{carbondensity}

\begin{figure}
    \centering
    \includegraphics[width=\linewidth]{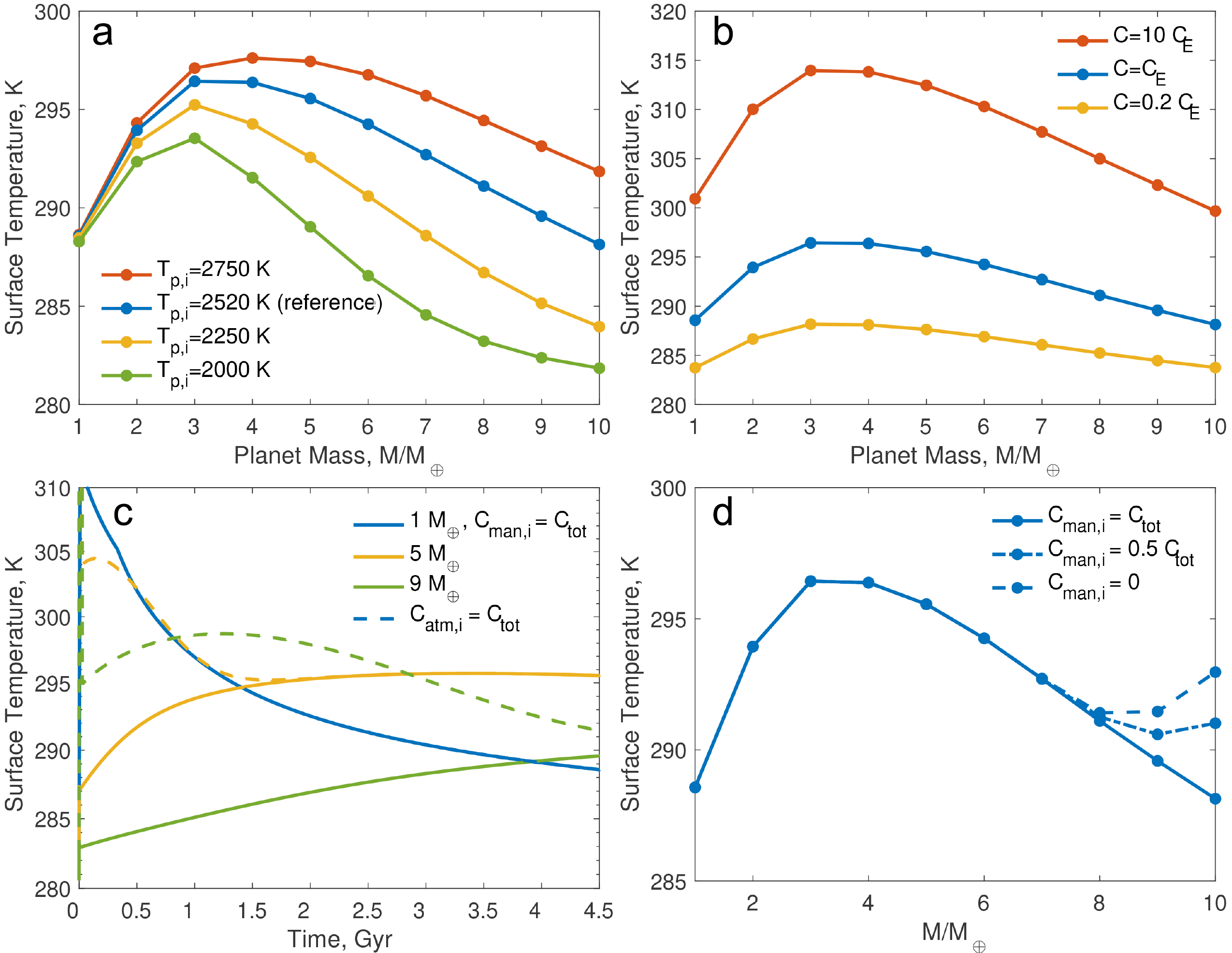}
    \caption{(a) and (b): Surface temperature at 4.5 Gyr as a function of planet mass for different (a) initial mantle temperatures and (b) bulk carbon densities $C$ relative to Earth's bulk carbon density $C_E$. Panel (c) compares the surface temperature evolution between planets with all carbon initially in the mantle (solid) and planets with all carbon initially in the atmosphere (dashed), and panel (d) depicts the surface temperature at 4.5 Gyr as a function of planet mass for different initial carbon distributions between the two reservoirs.}
    \label{fig:CDM}
\end{figure}

For Earth-sized planets the temperature-dependence of viscosity ensures that the late evolution hardly depends on the initial mantle temperature \citep{Schubert2001MantlePlanets}. This does not necessarily apply to more massive planets if the pressure-dependence of viscosity is considered \citep{Schaefer2015TheCycle}. For massive planets, the degassing rate at 4.5 Gyr and therefore the atmospheric CO$_2$ abundance increase with the initial mantle temperature, resulting in higher surface temperatures (see Fig. \ref{fig:CDM}a). However, the planet mass that yields the highest surface temperature at 4.5 Gyr is not very sensitive to the initial mantle temperature: In most cases, the peak remains at 3 $M_{\oplus}$, even though the initial mantle temperature is varied over a large range. However, for very high initial mantle temperatures, the peak is slightly shifted towards higher planet masses: For example, for a very high initial potential temperature of 2750 K, we find the peak at 4 $M_{\oplus}$ (Fig. \ref{fig:CDM}a).

Interestingly, numerical models of stagnant-lid planets \citep{Dorn2018OutgassingSuper-Earths} show a similar behavior, even though in these models the degassed CO$_2$ accumulates in the atmosphere over the entire evolution of the planet and is not only determined by the degassing rate at 4.5 Gyr. While \cite{Dorn2018OutgassingSuper-Earths} find a maximum degassed CO$_2$ for 3 $M_{\oplus}$ for an initial mantle temperature of 2000 K, this peak is shifted towards 2 $M_{\oplus}$ for an initially very low mantle temperature of 1600 K. Similarly, \cite{Noack2017VolcanismZone} find a peak in surface temperature at 4.5 Gyr for planets between 2 and 4 $M_{\oplus}$ with a shift of the peak with higher initial mantle temperatures towards higher planet masses. We note that the effect of the initial mantle temperature on the surface temperature at 4.5 Gyr is stronger for stagnant-lid than for plate tectonics planets, since the initial mantle temperature particularly controls early degassing, which makes up a significant part of the accumulated degassed CO$_2$ abundance on stagnant-lid planets in their late evolution.

The chemical composition of rocky exoplanets is thought to vary greatly \citep{bond2010compositional,Madhusudhan2016ExoplanetaryHabitability,moriarty2014chemistry}. Even for Earth it is still debated which combination of meteorites best represents Earth's bulk chemical composition \citep{Moynier2015TheBlocks}. In order to assess the effect of the planetary carbon density on the habitability of super-Earths, we ran models varying the carbon concentration and present the results in Fig. \ref{fig:CDM}b, keeping the land fraction constant at present-day Earth's value. Since the atmospheric CO$_2$ is controlled by the degassing rate, which in turn depends linearly on the mantle carbon concentration, our results do not change qualitatively. Similar to Fig. \ref{fig:CDM}a, we find a maximum surface temperature at 4.5 Gyr for planets of 3 Earth masses. The effect of planetary mass becomes more pertinent for higher bulk carbon concentrations, which has important implications for assessing the habitability of exoplanets: \cite{Kopparapu2013HABITABLEESTIMATES} argues that around 340 K the moist-greenhouse effect is reached, where the stratosphere becomes water-dominated and hydrogen loss to space occurs. This state can be considered as the inner edge of the habitable zone. For carbon-rich planets near the inner edge of the habitable zone, it is therefore crucial to consider the planetary mass when assessing the potential for life.

Fig. \ref{fig:CDM}c shows the effect of the initial carbon distribution between the mantle and the atmosphere on the evolution of the surface temperature (solid: all carbon initially in the mantle; dashed: all carbon initially in the atmosphere). While for planets of 1 Earth mass (blue) the initial carbon distribution is negligible, more massive planets need a longer period of time to redistribute their carbon: A planet of 5 Earth masses eliminates the effect of the initial distribution after 1.5 Gyr and a planet of 9 Earth masses is still slightly affected by the initial carbon distribution at 4.5 Gyr. This trend is supported by Fig. \ref{fig:CDM}d, which shows the surface temperature at 4.5 Gyr as a function of planet mass with all carbon initially in the mantle (solid), atmosphere (dashed), and equally distributed between the two reservoirs (dashed-dotted).

\section{Discussion}
\label{discussion}

Coupling a model of the long-term carbon cycle to thermal evolution models with temperature- and pressure-dependent viscosity and to a model of the melting depth, we assessed the influence of planetary mass on the surface temperature-evolution of Earth-like planets. In this section, we compare our model to previous studies, discuss simplifications and limitations of our model, and consequences for the habitability of exoplanets.

Starting with \cite{Walker1981ATemperature}, models of the long-term carbon cycle for Earth and other planets have been used to assess the climate evolution on Earth and other plate tectonics planets \citep{Sleep2001CarbonEarth,Kasting2003EvolutionPlanet,Foley2015ThePlanets}. A key parameter that influences the climate on long timescales is the mantle degassing rate, which is balanced by the weathering rate for specific values of atmospheric CO$_2$ and surface temperature. For Earth's evolution, it is suitable to approximate the degassing rate over time with scaling laws \citep{krissansen2017constraining,krissansen2018constraining,kadoya2020probable}, but in particular an extrapolation of the model to planets of different masses requires the consideration of the interior evolution.

\cite{Rushby2018Long-TermCycle} modelled the carbon cycle for planets of different masses by setting the degassing rate proportional to the internal heat production rate, thereby neglecting effects of the planet mass on the mantle viscosity and melting depth. As a result, they find only a little influence of planet mass on surface temperature, with a small trend towards higher surface temperature with increasing mass ($<$3 K difference between 1 and 10 Earth masses). However, this simplification neglects the fact that the melting depth below mid-ocean ridges crucially depends on the pressure-temperature profile and that the mantle heat flow (which controls the plate velocity) is determined by pressure- and temperature-dependent viscosity. In contrast, we find a peak in the surface temperature for 3 Earth mass planets, which is $\approx$8 K above the surface temperature for a 1 Earth mass planet. At higher masses, the surface temperature at 4.5 Gyr decreases again and may even fall below the surface temperature that is found for a 1 Earth mass planet. Whereas planets of 1 Earth mass have their highest mantle temperature and therefore degassing rate throughout the first 1 Gyr, after which the initial mantle temperature becomes unimportant (Schubert et al., 2001), the mantle temperature of more massive planets is much longer influenced by its initial value. Regardless of that, at 4.5 Gyr we find a peak in the surface temperature that is hardly influenced by initial conditions.

\cite{kadoya2015evolutionary} include a thermal evolution model of the planetary interior and calculate the degassing rate as a function of melting depth and plate velocity for planets up to 5 Earth masses. These authors use a viscosity law that neglects a pressure-dependence, which results (a) in a steadily increasing degassing rate with planetary mass and (b) in a steadily decreasing degassing rate with time for all planetary masses. These results are in accordance with \cite{oosterloo2021}, who calculate the plate velocity from a thermal evolution model of the mantle and derived the surface temperature evolution for different compositions and masses. In contrast, our results indicate that these trends do not hold if a pressure-dependence of viscosity is considered. In addition, we find that including a pressure limit to the melting depth substantially reduces the melting depth for massive planets.

The peak in surface temperature at 4.5 Gyr for planets of $\approx3$ Earth masses is similar to what has been found for stagnant-lid planets using numerical models of mantle convection in combination with a parameterization of the degassing rate \citep{Noack2017VolcanismZone,Dorn2018OutgassingSuper-Earths}. The reason for this is similar: First, the pressure-dependence of viscosity inhibits efficient heat flow and thereby high convection rates (or plate velocities in our model) for massive planets, whereas 1 Earth mass planets have already cooled down more strongly at 4.5 Gyr. Second, the increasing mantle temperature with planetary mass generally increases the melting depth up to the imposed pressure-limit to degassing, which reduces the effective melting depth for planets larger than 3 Earth masses.

Whether or not plate tectonics operate on exoplanets is a matter of debate and may depend on the plate thickness \citep{valencia2007Inevitability}, plate buoyancy \citep{Kite2009GeodynamicsPlanets}, and temperature \citep{Foley2012TheDamage,Noack2014PlateRheology}. Even for Earth, a transition from a stagnant-lid tectonic regime to modern plate tectonics may have occurred $\approx3$ billion years ago \citep{gerya2014precambrian,naeraa2012hafnium}. We note, however, that the surface temperature at 4.5 Gyr is hardly influenced by the nature of the early tectonic regime, as the atmospheric CO$_2$ is controlled by an equilibrium between degassing and weathering, with a residence time of CO$_2$ in the atmosphere that is far shorter than the timescale of interest here. Nevertheless, the influence of planetary mass on the likelihood of plate tectonics is crucial as it determines the applicability of our model. Plate tectonics is promoted by the localization of shear and the weakening of a lithosphere with a high viscosity. Weakening of the lithosphere can be caused by the formation of micro-cracks, defects or the reduction of grain-sizes \citep{Bercovici2014PlateInheritance}. Damage theory specifies this weakening of material. The damage mechanism described in \citet{Foley2012TheDamage} is a feedback system between grain-size reduction by deformation and a grain-size dependent viscosity: deformation reduces grain sizes, lowering the viscosity, resulting in more deformation. Larger planets exhibit lower healing of the lithosphere and higher convective stresses, and subsequently both factors increase the likelihood of plate tectonics to occur. Higher surface temperatures would promote the healing of the lithosphere by grain-growth. Therefore, \citet{Foley2012TheDamage} concludes that plate tectonics is more likely to occur on larger planets, preferably with cooler surface temperatures. Modelling results by \citet{Noack2014PlateRheology} indicate a peak likelihood for plate tectonics for planet masses between $M = 1$ and $M = 5$, suggesting that the effect of pressure on viscosity has a large influence on the occurrence of plate tectonics. Altogether, plate tectonics on planets somewhat more massive than Earth seems to be likely, although further studies are certainly needed.

An additional simplification of our model is the assumption of a constant land area. However, \cite{Abbot2012IndicationFraction} argue that lower land fractions should be expected for the more massive terrestrial planets. A planet's mass increases stronger than its surface area, which results in deeper oceans. The higher gravity furthermore reduces topography and creates shallower ocean basins \citep{Cowan2014WaterWaterworlds}, increasing the surface area covered with water.

Considering stellar evolution (Fig. \ref{fig:surfT}b), the temperatures found in the early evolution for massive planets are slightly below the freezing point of water. However, this does not necessarily imply a snowball state, since these temperatures should be regarded as globally averaged values and do not rule out an equatorial section with temperatures above freezing. According to \cite{Warren2002SnowballOcean}, at surface temperatures above 261 K ice layers are too thin ($<$1 m) to form coherent sheets. This point is supported by \cite{Feulner2014ClimateGlaciations} who find that the lowest steady-state configuration for a partially ice-covered state in the Sturnian glaciation generates a global mean surface air temperature of 266 K. For the Marinoan glaciation the coldest temperature at which ice-free ocean regions still exist is 268 K. In fact, \cite{Ye2015TheEarth} found preserved benthic macroalgae in the Marinoan-age Nantuo Formation in South China. According to the authors, this suggests that during glaciation muddy substrates existed in the photic zones, suggesting open-water areas in coastal environments. These areas might have acted as a shelter for benthic macro algae. Together with photoautotrophs, macro algae could have been part  of coastal food webs and the carbon cycle, just as they are today in glacial environments. In addition, autotrophes that do not rely on solar energy could exist below an icy crust. Altogether, life as we know it on Earth could potentially still evolve under these conditions. Nevertheless, as discussed above, massive planets are expected to keep their internal heat longer than planets of 1 Earth mass (see Fig \ref{fig:degassing}a). As a consequence, the degassing rate of massive planets remains high for a longer period of time, or even increases with time (see Fig \ref{fig:degassing}d). In combination with increasing stellar luminosity, the surface temperature of a massive planet that is in the conservative habitable range in the early evolution would increase stronger with time than that of a 1 Earth mass planet. We note that the initial mantle temperature affects the early degassing rate and could therefore also affect the onset time of habitability, but such a shift in time would not significantly affect the total duration of the habitable period. Altogether, with increasing mass of the planet, the time window for life to evolve on it appears to be become smaller.

\clearpage

\section{Conclusions}
In this paper, we have coupled a model of the long-term carbon cycle to a thermal evolution model of the mantle with temperature- and pressure-dependent viscosity and applied it to different planet masses. In particular, we calculated the degassing rate dependent on the melting depth and plate velocity. Our conclusions are summarized below:

\begin{itemize}
  \item The planet mass plays an important role in the evolution of the degassing rate, which causes different points in time throughout their evolution where the atmospheric CO$_2$ reaches its maximum. While low-mass planets have a high atmospheric CO$_2$ content particularly in their early evolution, high-mass planets are expected to build up CO$_2$-rich atmospheres later.
  \item At 4.5 Gyr, surface temperature increases with planetary mass up to $\approx$3 Earth masses, since (i) the temperature-dependence of mantle viscosity dominates over its pressure-dependence, which yields higher plate velocities with increasing planetary mass, and (ii) the melting depth increases with the mantle temperature. However, above $\approx$3 Earth masses, the surface temperature decreases with planetary mass, since (i) the pressure-dependence of viscosity dominates and (ii) a pressure-limit above which melt is not positively buoyant reduces the effective melting depth.
  \item Despite these effects of planetary mass, the long-term carbon cycle remains an important stabilizing feedback mechanism for massive Earth-like planets for a wide range of bulk carbon densities.
  \end{itemize}

\acknowledgments
We thank two anonymous reviewers for constructive comments, which greatly helped to improved the manuscript. DH has been supported through the NWA StartImpuls.

\appendix
\section{Carbon Cycle Model}
\label{carboncycle}
Seafloor weathering (section \ref{seafloorw}), continental weathering (section \ref{continentalw}), and carbonate subduction (section \ref{carbonatesubd}) follow from standard carbon cycle models \citep{Sleep2001CarbonEarth,Foley2015ThePlanets}. Parameters that directly depend on the planet size, such as land area or subduction zone length, are scaled accordingly.
\subsection{Seafloor Weathering Flux}
\label{seafloorw}
Seafloor weathering adds carbon to the oceanic plates, resulting in the withdrawal of CO$_\text{2}$ from the combined atmosphere and ocean reservoir. The CO$_\text{2}$ dissolved in seawater hydrothermally alters ocean floor basalt. Exactly how and to what extent the seafloor weathering flux operates is not fully understood at present. \cite{Brady1997SeafloorClimate} showed that the CO$_\text{2}$ uptake during hydrothermal alteration directly depends on the partial pressure of CO$_2$ in the atmosphere, $P_{CO_2}$. In addition, seafloor weathering demands the supply of fresh rock, which is created at mid-ocean ridges. The plate velocity therefore dictates the supply of weatherable material. These parameters are combined in the following equation as given by \cite{Foley2015ThePlanets}:
\begin{linenomath}
\begin{equation}
    F_{sfw} = F_{sfw}^*\Big(\frac{v_p}{v_{\oplus}}\Big)\bigg(\frac{P_{CO_2}}{P_{CO_2}^*}\bigg)^\alpha ,
\end{equation}
\end{linenomath}
where $v_{\oplus}$ is the modern day plate speed on Earth. The plate speed $v_p$ is derived in equation \ref{convvel}, and as ratio to $v_{\oplus}$ depicts the effect of spreading rate. This makes the seafloor weathering flux mass dependent, where increasing planetary size results in a higher $F_{sfw}$. $P_{\text{CO}_2}^*$ is the partial pressure for Earth's conditions and $\alpha = 0.25$. The present-day value for the seafloor weathering flux is $1.75 \cdot 10^{12}$ mol yr$^\text{-1}$ as determined by \cite{Mills2014ProterozoicWeathering}. Other scalings of seafloor weathering exist: \cite{krissansen2017constraining} argue for a direct temperature dependence of basalt dissolution and therefore of seafloor weathering. The effect of the applied scaling on our results is expected to be small, since continental weathering (section \ref{continentalw}) is the dominating flux controlling the climate, and directly depends on both, temperature and atmospheric CO$_2$.

\subsection{Continental Weathering Flux}
\label{continentalw}
The continental weathering flux $F_w$ is calculated following \cite{Foley2015ThePlanets} as follows:
\begin{linenomath}
\begin{equation} \label{eq:F_w}
    F_w = F_{ws} \left(1-\exp\Big[-\frac{f_l F_{w}^*}{f_l^* F_{ws}}\left(\frac{P_{CO_2}}{P_{CO_2}^*}\Big)^{\beta}\Big(\frac{P_{sat}}{P_{sat^*}}\right)^a \exp\Big(\frac{E_a}{R_g}\left(\frac{1}{T^*}-\frac{1}{T}\Big)\right)\Big]\right),
\end{equation}
\end{linenomath}
with the asterisk indicating values at present, $f_l$ denoting the land fraction, $E_a$ the activation energy, and $R_g$ the universal gas constant. The exponents $a$ and $\beta$ are constants. A breakdown of the remaining parameters is provided in the following.

A kinetically-limited continental weathering regime is assumed in this model. A kinetically controlled regime, although ill-constrained, is suggested for the Earth \citep{West2005TectonicWeathering}, indicating that the supply of weatherable bedrock exceeds the physical erosion rates necessary for complete denudation. In this regime, the parameter $F_{ws}$ calculates the supply limit to weathering and is retrieved by \cite{Foley2015ThePlanets} as follows:
\begin{linenomath}
\begin{equation}
    F_{ws} = A_P \frac{f_{l} E_r k_{cc}}{m_{cc}} \rho_r ,
\end{equation}
\end{linenomath}
where $f_{l}$ is the fraction of land above sea level that is subject to surface weathering, $A_P$ the surface area of the planet, $E_r$ is the erosion rate by physical processes, $k_{cc}$ the fraction of the cations Mg, Ca, K and Na in the continental crust, $\rho_r$ the regolith density and $m_{cc}$ the molar mass of the aforementioned cations.

With the supply limit to continental weathering established, three physical parameters remain to be determined for the weathering flux: the partial pressure of atmospheric CO$_\text{2}$, the temperature evolution of the atmosphere over time, and the saturation vapor pressure.

The partial pressure of atmospheric CO$_\text{2}$ can be related to the amount of carbon present in the atmosphere, $R_{atm}$. The initial value for $R_{atm}$ is assumed to be zero and CO$_\text{2}$ is added to the atmosphere by the degassing flux. The partial pressure affects the weathering rate by increasing the reaction rate with increasing PCO$_\text{2}$, and can be calculated by \citep{Foley2015ThePlanets}:
\begin{linenomath}
\begin{equation} \label{eq:PCO2}
    P_{CO_2} = R_{atm}\frac{mCO_2}{A_P}g_P , 
\end{equation}
\end{linenomath}
where $mCO_2$ is the molar mass of CO$_\text{2}$.

The surface temperature can be derived by using the effective temperature $T_e$. The effective temperature depends on the amount of solar irradiation and the albedo:
\begin{linenomath}
\begin{equation}
    T_e = \Big(\frac{S(1-A)}{4\sigma}\Big)^{1/4} ,
\end{equation}
\end{linenomath}
with $S$ the solar irradiance, $A$ the albedo of the planet, and $\sigma$ the Stefan Boltzmann constant. For the model that considers stellar evolution, $S$ is assumed to increase linearly by a factor of 1/3 throughout 4.5 Gyr \citep{ribas2009sun}.
The surface temperature follows from the effective temperature \citep{Foley2015ThePlanets}:
\begin{linenomath}
\begin{equation} \label{eq:T}
    T = T^*+2(T_e-T_e^*)+4.6\Big(\frac{P_{CO_2}}{P_{CO_2}^*}\Big)^{0.346}-4.6, 
\end{equation}
\end{linenomath}
where $T^*$, $T_e^*$, and $P_{{CO}_2}^*$ are the surface temperature, effective temperature, and CO$_2$ partial pressure of present-day Earth. Finally, the saturation vapor pressure $P_{sat}$ represents the fluctuation in runoff with changing surface temperature and can be calculated as follows \citep{Foley2015ThePlanets}
\begin{linenomath}
\begin{equation}
    P_{sat} = P_{sat0}\exp\Big[-\frac{m_wL_w}{R_g}\Big(\frac{1}{T}-\frac{1}{T_{sat0}}\Big)\Big], 
\end{equation}
\end{linenomath}
with $m_w$ denoting the molar mass of water, $L_w$ the latent heat of water, $P_{sat0}$ the reference saturation vapor pressure and $T_{sat0}$ the reference saturation vapor temperature.

\subsection{Subduction Flux}
\label{carbonatesubd}
The concluding part of the long-term carbon cycle in this model is the subduction flux. Unlike other atmospheric gases, CO$_\text{2}$ reacts with water to form carbonic acid rain, dissolving silicate rock through weathering. Carbon is then removed from the atmospheric system by chemical reaction of the dissolved CO$_\text{2}$ with Ca and Mg containing silicate minerals. This can be in the form of sediment as limestone (CaCO$_\text{3}$), or as organic matter (CH$_\text{2}$O), and is emplaced on the oceanic plates. The fate of 99\% of the oceanic plates is to be subducted. Part of the carbon content on the plates will be accreted to the overriding plate, but it is estimated that this term is approximately balanced by the addition of carbon to the subducting plate, through tectonic erosion \citep{Sleep2001CarbonEarth}. At any time, the amount of carbon $R_{p}$ on the oceanic plates can be calculated by the mass balance equation:
\begin{linenomath}
\begin{equation} \label{eq:Rp}
    \frac{dR_p}{dt} = \frac{F_w}{2}-F_{sub} + F_{sfw}.
\end{equation}
\end{linenomath}
The weathering flux $F_w$ as obtained by equation \ref{eq:F_w} is divided by two to account for the carbon that is re-released to the atmosphere when the carbonates are formed. Then, the total amount of subducted carbon can be obtained by \citep{Foley2015ThePlanets}:
\begin{linenomath}
\begin{equation}
    F_{sub} = \frac{R_p L}{A_{plate}}v_p ,
\end{equation}
\end{linenomath}
where $A_{plate}$ represents the area of the ocean plates. The area of the plates is retrieved by multiplying $(1-l_f)$ with the surface area of a planet. The parameters $v_{p}$ and $L$ are the velocity of the plates and length of trenches, respectively.

It needs to be accounted for that not all of the subducted carbon will reach the deeper parts of the mantle, but instead that part of it will be re-emitted into the atmosphere by arc volcanism. This is achieved by establishing the arc volcanism flux $F_{arc}$:
\begin{linenomath}
\begin{equation}
    F_{arc} = f F_{sub} , 
\end{equation}
\end{linenomath}
where $f$ is the fraction that degasses by arc volcanism. As of today, $f$ is not well-constrained and estimates vary as much as 25-70\%
\citep{Foley2015ThePlanets}. Hence, an average value of 50\% will be assumed since it does not affect the outcomes in significant ways. 

Finally, the reservoirs need to be connected through the fluxes. The amount of carbon on oceanic plates was earlier established in equation \ref{eq:Rp}, so the mantle and atmospheric reservoirs are left to be determined. The quantity of carbon in the atmosphere and oceans combined is given by the mass balance equation \citep{Foley2015ThePlanets}:
\begin{linenomath}
\begin{equation}
    \frac{dR_{atm}}{dt} = F_{deg} + F_{arc} - \frac{F_w}{2} - F_{sfw},
\end{equation}
\end{linenomath}
and the amount of carbon in the mantle at any given time is:
\begin{linenomath}
\begin{equation}
    \frac{dR_{man}}{dt} = (1 - f)F_{sub} - F_{deg}.
\end{equation}
\end{linenomath}

\begin{table}
\centering
 \caption{Parameter values used for the carbon cycle model, adopted from \cite{Foley2015ThePlanets}}
 \label{table:FoleyParameters}
 \begin{tabular}{l l l l} 
 \textbf{Parameter} & \textbf{Definition} & \textbf{Value} \\ [0.5ex] 
 \hline

$f_l$ & land fracion & $0.3$ \\
$E_r$ & maximum erosion rate & $0.01$ m yr$^\text{-1}$ \\
$\rho_{r}$ & density of regolith & 2500 kg m$^\text{3}$ \\
$m_{cc}$ & average molar mass Mg,Ca,K,Na & 0.032 kg m$^\text{3}$ \\
$k_{cc}$ & fraction Mg,Ca,K,Na in crust & $0.08$ \\
$S(4.5 Gyr)$ & present-day solar irradiance & 1373 W m$^\text{-2}$ \\
$A$ & albedo & $0.31$ \\
$\sigma$ & Stefan Boltzmann constant & $5.67\cdot 10^{-8}$ W m$^\text{-2}$   \\
$mCO_2$ & molar mass CO$_\text{2}$ & 0.044 kg mol$^\text{-1}$ \\
$T^*$ & present day temperature & 285 K \\
$T_e^*$ & present day effective temperature & 254 K \\
$PCO_2^*$ & present day partial pressure CO$_\text{2}$ & 33 Pa \\
$F_w^*$ & present day weathering flux & $12\cdot10^{12}$ mol yr$^\text{-1}$ \\
$P_{sat0}$ & reference saturation vapor pressure & 610 Pa \\
$T_{sat0}$ & reference temperature & 273 K \\
$E_a$ & activation energy silicate weathering & $42\cdot 10^3$ J mol $^\text{-1}$ \\
$a$ & $P_{sat}$ exponent for silicate weathering & 0.3 \\
$\alpha$ & PCO$_\text{2}$ exponent for seafloor weathering & $0.25$ \\
$\beta$ & PCO$_\text{2}$ exponent for silicate weathering & $0.55$ \\
$P_{sat}^*$ & present day value saturation vapor pressure & 1691 Pa \\
$f$ & fraction of subducted carbon that degasses & $0.5$\\
$F_{sfw}^*$ & present day seafloor weathering flux & $1.75\cdot10^{12}$ mol yr$^\text{-1}$\\
\hline
\end{tabular}
\end{table}

\section{Thermal Evolution Model}
\label{thermalevolution}

The spherically averaged mantle temperature $\langle T_m \rangle$ evolves as
\begin{linenomath}
\begin{equation}
\label{heattransfer}
	\rho_m C_p V_m \frac{d \langle T_m \rangle}{dt} = -A_s q_s + A_c q_c+V_m Q(t),
\end{equation}
\end{linenomath}
where $\rho_m$, $C_p$, $V_m$ are the mantle density, heat capacity, and volume, $A_s$ and $A_c$ are the surface area of the planet and the core, and $q_s$ and $q_c$ are the surface and core heat flux, respectively. For the heat production $Q_m(t)$ we assume Earth's relative mantle abundances of the main radiogenic heat producing elements K, U and Th, as given by \citep{Korenaga2008UreyMantle}:
\begin{linenomath}
\begin{equation}
\label{heatprod}
Q_m(t)=Q_0\sum_i q_i \exp\left\lbrace-\frac{\log(2)}{h_i}(t-4.5)\right\rbrace,
\end{equation}
\end{linenomath}
where $t$ is the time in Gyr, $Q_0$ is the total present-day heat production rate, and  $q_1...q_4$ and $h_1...h_4$ are the relative present-day heat production rates and half-life times of $^{238}$U, $^{235}$U, $^{232}$Th, and $^{40}$K as given in \cite{Korenaga2008UreyMantle}.

The mantle heat flux is given by
\begin{linenomath}
\begin{equation}
\label{hfmantle}
	q_m = k \frac{(T_u-T_s)}{\delta_u},
\end{equation}
\end{linenomath}
where $T_u$ is the temperature at the base of the upper boundary layer (the crust-mantle boundary), $T_s$ the surface temperature.

The core temperature evolution is calculated based on the assumption that there are negligible concentrations of K, U and Th in the core, and is given by
\begin{linenomath}
\begin{equation}
\label{dtcore}
    \rho_cC_{p,c}V_c\frac{dT_c}{dt} = -A_cq_c.
\end{equation}
\end{linenomath}

\bibliography{references}

\end{document}